\documentclass[acmsmall,nonacm]{acmart}

\usepackage{tikz}
\usepackage[linesnumbered,ruled,vlined]{algorithm2e}

\AtBeginDocument{%
  \providecommand\BibTeX{{%
    \normalfont B\kern-0.5em{\scshape i\kern-0.25em b}\kern-0.8em\TeX}}}

\setcopyright{acmcopyright}
\copyrightyear{2020}
\acmYear{2020}
\acmDOI{10.1145/1122445.1122456}

\acmJournal{TOCT}
\acmVolume{1}
\acmNumber{1}
\acmArticle{1}
\acmMonth{1}

\begin{document}

\title{Signal filtering to obtain number of Hamiltonian paths}

\author{Bryce Kim}
\email{brycemkim@gmail.com}

\begin{abstract}
This paper consists of two parts. First, the (undirected) Hamiltonian path problem is reduced to a signal filtering problem - number of Hamiltonian paths becomes amplitude at zero frequency for (a combination of) sinusoidal signal f(t) that encodes a graph. Then a 'divide and conquer' strategy to filtering out wide bandwidth components of a signal is suggested - one filters out angular frequency 1/2 to 1, then 1/4 to 1/2, then 1/8 to 1/4 and so on. An actual implementation of this strategy involves careful local polynomial extrapolation using numerical differentiation filters. When conjectures regarding required number of samples for specified filter designs and time complexity of obtaining filter coefficients hold, P=NP conditionally.
\end{abstract}



\maketitle

\section{Introduction}
\label{sec:intro}
This paper consists of two parts (1) and (2):
\begin{enumerate}
\item function encoding of an undirected graph into $f(t)$ such that the zero-frequency amplitude $A_0$ of $f(t)$ is the number of Hamiltonian paths: $n_h$. Furthermore, Algorithm \ref{alg:algxt} computes $x(t)$ in polynomial time relative to $|V|\equiv \eta$ for each $t$. 
\item lowpass wide bandwidth filtering strategy, consisting of two sub-strategies: frequency binary divide-and-conquer and numerical differentiation filter-based local polynomial extrapolation sub-strategies.
\item Filtering out $f(t)$ to obtain its zero frequency amplitude takes polynomial time relative $\eta$, when conjectures of Equation \eqref{eqn:filterordernfnd}, Equation \eqref{eqn:filcoeffbound} and Equation \eqref{eqn:ffctimecomplexity}, stating the minimal required filter order bound, the filter coefficient magnitude upper bound and the filter coefficient time complexity bound, hold. This part dominates time complexity, which one can state as $O(\eta^{k_{ffc}})$, with $k_{ffc}$ defined in Equation \eqref{eqn:kffc}.
\end{enumerate}

\begin{itemize}
\item Corresponding to above (1): a way of encoding an undirected graph $G=(V,E)$ (with $|V|=\eta$) into a combination $f(t)$ of sinusoidal signals, with $f(t) \in \mathbb{C}$ and $t \in \mathbb{R}$, is shown. In frequency domain $F(\omega)$ of $f(t)$, $F(0)$ turns out to be the number of Hamiltonian paths. Therefore, finding the number of Hamiltonian paths is reduced to finding $F(0)$ by lowpass filtering.\newline

This is done by assigning each vertex (with assigned ordinal $i$ when vertices are ordered from 1 to $\eta$) angular frequency of $v_i = \eta^i$ appropriately. Let a sum of vertices refers to a sum of these vertex angular frequencies. Then each $\eta$-walk $w$ - walk of length $\eta-1$ allowed in $G$ - is assigned angular frequency $\omega_w$ that is the sum of all vertices visited. For example, if $v_1$ is visited twice and $v_2$ is visited once in 3-walk $w$, then $\omega_w = 2v_1 + v_2$. The goal then is to assign $w$ with signal $e^{i\omega_w t}$, and then sum up all possible $\eta$-walk signals to form $x(t) = \sum_w e^{i\omega_w t}$. The angular frequency of Hamiltonian paths in $x(t)$ then would be $\eta_h = \sum_i v_i$. To make signal filtering convenient, we shift the Hamiltonian path angular frequency from $\eta_h$ to zero by $y(t) = x(t)e^{-i\eta_h t}$. Then we re-scale time by $f(t) = y(t/\eta^{\eta+1})$.\newline

The remaining issue then is to demonstrate that the Hamiltonian path angular frequency is not shared by other paths - this is done by utilizing the basis representation theorem, with basis being $|V| \equiv \eta$ and vertex frequencies assigned as $V = \{\eta,\eta^2,..,\eta^{\eta}\}$, given that each $\eta$-walk (a walk of length $\eta-1$) cannot visit the same vertex $\eta$ times.\newline

The rest of the proof that Algorithm \ref{alg:algxt}, which computes $x(t)$, works as advertised is given in subsection \ref{subsec:explainalgxt}.

\item Corresponding to above (2): the problem with $f(t)$ turns out to be that non-zero angular frequencies of $f(t)$ with potential non-zero amplitude have domain of $1/\eta^{\eta+1} \leq |\omega| \leq 1$. That is, the least positive angular frequency with potential non-zero amplitude is $1/{\eta}^{\eta+1}$, with maximum angular frequency with potential non-zero amplitude being $1$. Conventionally, this filtering problem is considered to be infeasible, requiring exponential computational resource (relative to $\eta$). This paper presents a `divide and conquer' strategy that demonstrates polynomial computational resource, in sense that time complexity is $O(\eta^{k_{ffc}})$, where $k_{ffc}$ is some constant, assuming a particular digital filter order upper bound when given filter design constraints and polynomial time complexity relative to $\eta$ in obtaining filter coefficients. In that sense $P=NP$ up the filter order bound and coefficient time complexity assumptions.\newline

This is done by effectively first filtering out 1/2 of active angular frequencies, extrapolate new samples from filter output samples, and then filtering out 1/4 of originally active angular frequencies, extrapolate new samples and then filtering out 1/8 of originally active angular frequencies and so forth. This allows us the continued use of the same modest cutoff frequency, instead of an extremely low cutoff frequency that $f(t)$ seems to demand for.\newline

Extrapolation error is controlled by re-doing numerical differentiation whenever new samples are extrapolated using a constructed local polynomial that utilize numerical differentiation data - this allows us to treat extrapolation errors as if they come from a combination of sinusoid inputs, given that numerical differentiation filters admit the frequency response interpretation. This allows us to tame down extrapolation errors. Error analysis then is conducted with heavy use of Parseval's theorem and by the `extrapolation error of extrapolation error' strategy.
\end{itemize}

\subsection{Preliminary assumption}
Sufficiently large $|V| \equiv \eta > \eta_0$ would be assumed throughout the paper for simplification purposes.

\subsection{Notation style}
For notation simplicity, it would be assumed that $\eta^{\eta^x} \equiv \eta^{\left(\eta^x\right)}$. $j$, in contrast to typical engineering convention, would not refer to an imaginary number. Depending on contexts, $i$ would either be used as an imaginary number or index $i$, as typical in mathematical convention. $k$ would refer to positive natural number constants, with different subscripts. For notation, $a^{b^c} \equiv a^{\left(b^c\right)} \neq \left(a^b\right)^c$. By cycles, they would not refer to graph-theoretic cycles or Hamiltonian cycles and would be defined differently.

For formula that are technically not equations, they may still be referred to as equations for expositional convenience. Superscripts always represent exponentiation. Sampling interval is always assumed to be $\Delta t = 1$. $z \equiv N_f+N_d+1$, $z_2 \equiv 2(N_f+N_d)+N_f-2$. Time refers to variable $t$ in functions such as $f(t)$.

\subsection{Preliminary terminology}
A `sinusoid(al) contribution' would always refer to a combination of sinusoid(al) contributions. That is, (a combination of) sinusoid(al) contributions, where the term in the parenthesis would often be not written.

From here on, each vertex would be labelled with its angular frequency, instead of its ordinal based on order from 1 to $\eta$, as $V = \{\eta,\eta^2,..,\eta^{\eta}\}$ shows. Each vertex with assigned ordinal $i$ is assigned vertex label and angular frequency of $\eta^i$. A sum of vertices then refers to a sum of vertex angular frequencies (labels).

In contrast to vertices, a sum $q(t)$ of walks refers to a sum of signals with walk angular frequencies. Or formally notated, $q(t) = \sum_{w}e^{i\omega_{w,q}}t$, where $w$ refers to each walk being summed up to form $q(t)$, and $\omega_{w,q}$ refers to angular frequency of walk $w$.

A $\eta$-walk refer to a walk of length $\eta-1$, with restrictions that edges connecting one vertex to itself are disallowed and that a walk has to respect allowed edge connections. Edges may be used more than once. Length of a walk refers to the number of times any edge is walked upon by the walk. Or equivalently, it refers to the number of times any vertex is visited minus one. Each walk can visit one vertex more than once, as far as edge connections allow it. 

\subsection{Style issues in exponents and subscripts}
Due to the style of the manuscript, exponents and superscripts may not be clearly identified. $\theta_c$ in $g_{\theta_c}(t)$, $w_{\theta_c,j}(t)$, $\gamma_{\theta_c}(t)$, $\psi_{\theta_c}(t)$, $\gamma_{\theta_c,j}(t)$, $\psi_{\theta_c,j}(t)$, $\upsilon_{\theta_c,j}(t)$ and $\gamma_{\theta_c,j}(t)$ of subsubsection \ref{subsubsec:errstrategy} are subscripts. The denominator of $\frac{2}{2^{\eta^{k_{baf}}}}$ should be read as two to the power of $\eta^{k_{baf}}$. Similarly, denominator $2^{\eta^{k_{ddd..}}}$ in $\frac{2\eta^{..}}{2^{\eta^{k_{ddd..}}}}$ should be read as two to the power of $\eta^{k_{ddd..}}$, where $ddd..$ refers to some random subscript. $2^{\eta^{k_{ddd..}}}$ must be distinguished with $2\eta^{..}$, which refers to 2 times $\eta^{..}$.

\section{Function encoding of graph}
\label{sec:funcencode}
The idea is to encode or translate an undirected graph as a computation circuit - for example, for graph in Figure \ref{fig:stang}, the circuit of Figure \ref{fig:expandg} is generated. We use the circuit to generate $x(t)$, from which we generate $f(t)$.

\subsection{What is x(t)?}
\label{subsec:xt}
Essentially, $x(t)$ is the sum of all $\eta$-walks, which refer to walks of length $\eta-1$. (Each $\eta$-walk contains $\eta$ vertices, potentially repeated, and $\eta-1$ edges, potentially repeated. This follows standard graph theory terminology.) Each $\eta$-walk $w$ is assigned angular frequency $\omega_w$ - thus each walk represents $e^{i\omega_w t}$. That is, $x(t) = \sum_{w} e^{i\omega_w t}$, where $w$ is restricted to a possible $\eta$-walk according to graph $G=(V,E)$.

The eventual goal of this paper is to calculate the number of Hamiltonian paths. Therefore, it is essential to distinguish Hamiltonian paths from other $\eta$-walks. One simple implementation is to assign a unique frequency to walks sharing same vertices up to permutation. That is, in Figure \ref{fig:stang}, one possible $4$-walk is A-B-A-C and another is C-A-B-A. These walks share the following profile: A has been visited twice, B has been visited once, C has been visited once, D has been visited zero times. This is the implementation followed here.

Then the issue is, 1) the exact way frequency is assigned to each walk, 2) how $x(t)$ then may efficiently be computed without having to compute each walk signal $w(t)$ and then summing up.

\subsection{Vertex and walk frequency assignment}
\label{subsec:vertexwalk}
One first orders vertices in $V$ and assigns ordinal number $i$ to each vertex, with $i$ ranging from $1$ to $\eta$. Then angular frequency is assigned to each vertex as $\eta^i$, which would directly be used to refer to the vertex.

Then a walk is assigned angular frequency as sum of vertices. That is, $\omega_w = \sum_v c_v v$,  where $v$ is a vertex visited by walk $w$ and $c_v$ is the number of times $v$ is visited by walk $w$. One can then see that walks with different vertex visit counts have distinct angular frequencies via the basis representation theorem - $\eta$ essentially works as basis, and one vertex can only be visited less than $\eta$ times: $V = \{\eta,\eta^2,..,\eta^{\eta}\}$.

\subsection{How x(t) and f(t) are computed}
\label{subsec:xtcompute}
Recall that $x(t)$ is sum of all $\eta$-walks. The computation procedure is stated in Algorithm \ref{alg:algxt}.
\begin{algorithm}
\SetAlgoLined
\SetKwInOut{Input}{input}
\SetKwInOut{Output}{output}
\Input{Graph $G=(V,E)$, $|V|=\eta$, $V=\{\eta,\eta^2,..,\eta^{\eta}\}$, time $t$}
\Output{$x(t)$}
 Initialize array $u_1$ and $u_2$ of length $\eta$, index starting from 1\;
 $x \gets 0$\;
 \For{$j \gets 1$ \KwTo $\eta$}{
    $u_1[j] \gets e^{i\eta^j t}$
 } 
 \For{$j \gets 2$ \KwTo $\eta$}{
    \For{$m \gets 1$ \KwTo $\eta$}
    {
        $u_2[m] \gets 0$\;
        \For{$p \gets 1$ \KwTo $\eta$}
        {
            \If{$(\eta^p, \eta^m) \in E$}
            {
                $u_2[m] \gets u_2[m] + u_1[p]$\;
            }   
        }
        $u_2[m] \gets u_2[m] e^{i\eta^m t}$\;
    }
    \For{$m \gets 1$ \KwTo $\eta$}
    {
        $u_1[m] \gets u_2[m]$\;    
    }
 }
 \For{$j \gets 1$ \KwTo $\eta$}
 {
    $x \gets x + u_1[j]$
 }
 
 \caption{computation of $x(t)$}
 \label{alg:algxt}
\end{algorithm}
Algorithm \ref{alg:algxt} may be represented in the expanded circuit form, which is shown in Figure \ref{fig:expandg} for the graph in Figure \ref{fig:stang}. Here we do not yet consider inevitable computation errors.

Multiplying $x(t)$ by $e^{-i\eta_h t}$, where $\eta_h = \sum_{i=1}^{\eta}{\eta}^i$ ($\eta_h$ is then the angular frequency of Hamiltonian paths in $x(t)$ by definition), gives us $y(t) = x(t)e^{-i\eta_h t}$. This moves the angular frequency of Hamiltonian paths to zero. Then time is re-scaled such that $f(t) = y(t/{\eta}^{\eta+1})$. This makes angular frequency with potential non-zero amplitude to range from $0$ to $1$ with $\Delta \omega = 1/{\eta}^{\eta +1}$.

Then it can be seen that finding the number of Hamiltonian paths is about finding $F(0)$ in frequency domain of $f(t)$, which means lowpass filtering of $f(t)$, just with wide bandwidth to be filtered out.

\begin{figure}[h]
\centering
\caption{A 4-vertex graph}
\label{fig:stang}
\begin{tikzpicture}
    \node[shape=circle,draw=black,align=left] (A) at (0,0) {A};
    \node[shape=circle,draw=black] (B) at (0,3) {B};
    \node[shape=circle,draw=black] (C) at (3,3) {C};
    \node[shape=circle,draw=black] (D) at (3,0) {D};

    \path [-] (A) edge node[left] {} (B);
    \path [-] (A) edge node[left] {} (C);
    \path [-](B) edge node[left] {} (C);
    \path [-](A) edge node[left] {} (D);
    \path [-](D) edge node[left] {} (C); 
\end{tikzpicture}
\end{figure}

\begin{figure}[h]
\centering
\caption{The circuit expansion representation of the graph in Figure \ref{fig:stang}. Edges in this figure are wires consistent with edge connections in the graph, with circles representing vertices at each level. $\mathtt{\sim}$ represents function generator associated with each vertex, $\times\mathtt{\sim}$ represents multiplying the sum coming from the adder $\sum$ with function generator (oscillator) associated with each vertex. L1, L2, L3, L4 represent levels. Function generator at each vertex circle with angular frequency $v$ sends out $e^{ivt}$ to every outgoing wire. Adder simply sums up values coming from incoming wires.}
\label{fig:expandg}
\begin{tikzpicture}[scale=0.7, every node/.style={scale=0.7}]
    \node[shape=circle,draw=black,align=center] (A1) at (0,6) {$\mathtt{\sim}$\\A};
    \node[shape=circle,draw=black,align=center] (B1) at (0,4) {$\mathtt{\sim}$\\B};
    \node[shape=circle,draw=black,align=center] (C1) at (0,2) {$\mathtt{\sim}$\\C};
    \node[shape=circle,draw=black,align=center] (D1) at (0,0) {$\mathtt{\sim}$\\D};
    \node[shape=rectangle,draw=black] (LY1) at (0,-2) {L1};

	\node[shape=circle,draw=black,align=center] (A2) at (3,6) {$\times \mathtt{\sim}$\\A};
	\node[shape=circle,draw=black,align=center] (B2) at (3,4) {$\times\mathtt{\sim}$\\B};
	\node[shape=circle,draw=black,align=center] (C2) at (3,2) {$\times\mathtt{\sim}$\\C};
	\node[shape=circle,draw=black,align=center] (D2) at (3,0) {$\times\mathtt{\sim}$\\D};
    \node[shape=rectangle,draw=black] (LY2) at (3,-2) {L2};

	\node[shape=circle,draw=black,align=center] (A3) at (6,6) {$\times \mathtt{\sim}$\\A};
	\node[shape=circle,draw=black,align=center] (B3) at (6,4) {$\times \mathtt{\sim}$\\B};
	\node[shape=circle,draw=black,align=center] (C3) at (6,2) {$\times \mathtt{\sim}$\\C};
	\node[shape=circle,draw=black,align=center] (D3) at (6,0) {$\times \mathtt{\sim}$\\D};
    \node[shape=rectangle,draw=black] (LY3) at (6,-2) {L3};

	\node[shape=circle,draw=black,align=center] (A4) at (9,6) {$\times \mathtt{\sim}$\\A};
	\node[shape=circle,draw=black,align=center] (B4) at (9,4) {$\times \mathtt{\sim}$\\B};
	\node[shape=circle,draw=black,align=center] (C4) at (9,2) {$\times \mathtt{\sim}$\\C};
	\node[shape=circle,draw=black,align=center] (D4) at (9,0) {$\times \mathtt{\sim}$\\D};
    \node[shape=rectangle,draw=black] (LY4) at (9,-2) {L4};

    \node[shape=circle,draw=black,align=center] (SA2) at (1.8,6) {$\sum$};
    \node[shape=circle,draw=black,align=center] (SB2) at (1.8,4) {$\sum$};
    \node[shape=circle,draw=black,align=center] (SC2) at (1.8,2) {$\sum$};
    \node[shape=circle,draw=black,align=center] (SD2) at (1.8,0) {$\sum$};

    \node[shape=circle,draw=black,align=center] (SA3) at (4.8,6) {$\sum$};
    \node[shape=circle,draw=black,align=center] (SB3) at (4.8,4) {$\sum$};
    \node[shape=circle,draw=black,align=center] (SC3) at (4.8,2) {$\sum$};
    \node[shape=circle,draw=black,align=center] (SD3) at (4.8,0) {$\sum$};

    \node[shape=circle,draw=black,align=center] (SA4) at (7.8,6) {$\sum$};
    \node[shape=circle,draw=black,align=center] (SB4) at (7.8,4) {$\sum$};
    \node[shape=circle,draw=black,align=center] (SC4) at (7.8,2) {$\sum$};
    \node[shape=circle,draw=black,align=center] (SD4) at (7.8,0) {$\sum$};

	\node[shape=circle,draw=black,align=center] (SF) at (10.3,3) {$\sum$};

    \node[align=center] (xt) at (11.3,3) {$x(t)$};

	\path [->] (A1) edge node[left] {} (SB2);
    \path [->] (A1) edge node[left] {} (SC2);
	\path [->] (A1) edge node[left] {} (SD2);
    \path [->] (B1) edge node[left] {} (SA2);
    \path [->] (B1) edge node[left] {} (SC2);
    \path [->] (C1) edge node[left] {} (SA2);
    \path [->] (C1) edge node[left] {} (SB2);
    \path [->] (C1) edge node[left] {} (SD2);
	\path [->] (D1) edge node[left] {} (SA2);
    \path [->] (D1) edge node[left] {} (SC2);

	\path [->] (SA2) edge node[left] {} (A2);
	\path [->] (SB2) edge node[left] {} (B2);
	\path [->] (SC2) edge node[left] {} (C2);
	\path [->] (SD2) edge node[left] {} (D2);

	\path [->] (A2) edge node[left] {} (SB3);
    \path [->] (A2) edge node[left] {} (SC3);
	\path [->] (A2) edge node[left] {} (SD3);
    \path [->] (B2) edge node[left] {} (SA3);
    \path [->] (B2) edge node[left] {} (SC3);
    \path [->] (C2) edge node[left] {} (SA3);
    \path [->] (C2) edge node[left] {} (SB3);
    \path [->] (C2) edge node[left] {} (SD3);
	\path [->] (D2) edge node[left] {} (SA3);
    \path [->] (D2) edge node[left] {} (SC3);

	\path [->] (SA3) edge node[left] {} (A3);
	\path [->] (SB3) edge node[left] {} (B3);
	\path [->] (SC3) edge node[left] {} (C3);
	\path [->] (SD3) edge node[left] {} (D3);

	\path [->] (A3) edge node[left] {} (SB4);
    \path [->] (A3) edge node[left] {} (SC4);
	\path [->] (A3) edge node[left] {} (SD4);
    \path [->] (B3) edge node[left] {} (SA4);
    \path [->] (B3) edge node[left] {} (SC4);
    \path [->] (C3) edge node[left] {} (SA4);
    \path [->] (C3) edge node[left] {} (SB4);
    \path [->] (C3) edge node[left] {} (SD4);
	\path [->] (D3) edge node[left] {} (SA4);
    \path [->] (D3) edge node[left] {} (SC4);

	\path [->] (SA4) edge node[left] {} (A4);
	\path [->] (SB4) edge node[left] {} (B4);
	\path [->] (SC4) edge node[left] {} (C4);
	\path [->] (SD4) edge node[left] {} (D4);

	\path [->] (A4) edge node[left] {} (SF);
	\path [->] (B4) edge node[left] {} (SF);
	\path [->] (C4) edge node[left] {} (SF);
	\path [->] (D4) edge node[left] {} (SF);

	\path [->] (SF) edge node[left] {} (xt);

\end{tikzpicture}
\end{figure}

\subsection{Explaining the x(t) algorithm}
\label{subsec:explainalgxt}
Let us prove that Algorithm \ref{alg:algxt} does what it is supposed to do - $x(t)$ as the sum of all $\eta$-walks, with vertex frequency $V=\{\eta,\eta^2,..,\eta^{\eta}\}$ by induction. 

Figure \ref{fig:expandg} helps in the proof - let us use level notations there. Then at the Line 9-13 loop of Algorithm \ref{alg:algxt}, it is at the vertex with index $m$ (or simply, notated with vertex angular frequency, $\eta^m$), with level denoted by index $j$ of Line 6 in the algorithm. In the Line 9-13 loop, it searches for all vertices $\eta^p$ connected to vertex $\eta^m$ by allowed edges and adds up values $u_1[p]$ stored in these vertices.

$u_1[p]$ (is assumed to and would be proven to) stands for the sum of all $j-1$-walks that end with vertex $p$. Therefore, summing up $u_1[p]$ for $(\eta^p,\eta^m)\in E$ stands for the sum of all $j-1$-walks that can be connected with vertex $\eta^m$ at the end to form $j$-walks. Then Line 14 of the algorithm multiplies the resulting sum $u_2[m]$ with $e^{i\eta^m t}$, which (is assumed to and would be proven to) makes $u_2[m]$ the sum of all $j$-walks that end with vertex $\eta^m$. Note that $j$ in $j$-walks match with level $j$ in Line 6 of the algorithm. Then at Line 16, value of $u_2[m]$ is copied to $u_1[m]$. 

Line 4 simply states that for 1-walks (level 1), $u_1[j]$ should be $e^{i\eta^j t}$. Line 21 states that after Level $\eta$, one needs to sum up all $\eta$-walks to form $x(t)$.

\subsubsection{The main proof of the x(t) algorithm}
Now onto proving that each $u_1[m]$ created at Line 17 of Algorithm \ref{alg:algxt} and level $j$ faithfully captures angular frequency of every $j$-walk ending with vertex $\eta^m$. The proof goes by induction.

Suppose that to level $j$, each $u_1[p]$ faithfully captures angular frequency (and associated signals) of every $j$-walk $w_p$ that ends with vertex $\eta^p$. Then $u_1[p](t) = \sum_{w_p} e^{i\omega_{w_p} t}$, where $\omega_{w_p}$ refers to its angular frequency. Then $u_2[m]$ at Line 17 of the algorithm is:
\begin{multline}
\label{eqn:u2ml17}
u_2[m](t) = \sum_{(\eta^p,\eta^m)\in E}u_1[p](t) =\\
\left[\sum_{(\eta^p,\eta^m)\in E}\sum_{w_p} e^{i\omega_{w_p} t}\right]e^{i\eta^m t} =\\
\left[\sum_{(\eta^p,\eta^m)\in E}\sum_{w_p} e^{i\left(\omega_{w_p} + \eta^m \right)t}\right]
\end{multline}
Equation \eqref{eqn:u2ml17} demonstrates that to existing $j$-walk angular frequency, $\eta^m$ was added to produce $j+1$-walk angular frequency, as demanded for walk angular frequency. Therefore, the inductive step from level $j$ to level $j+1$ was demonstrated. One thus just needs to demonstrate that at level 1, each $u_1[p]$ faithfully captures angular frequency of 1-walk. This is true by design at Line 4 of the algorithm. Therefore, the proof by induction is complete - $x(t)$ faithfully captures angular frequency (and associated signals) of every $\eta$-walk. 
 
\section{Lowpass wide bandwidth filtering}
\label{sec:lowpasswide}
\subsection{Section introduction}
For a conventional lowpass filter setup, if a required cutoff frequency is so low relative to maximum magnitude of input signal $f(t)$, then too many samples are required for successful lowpass filtering. Let us provide the example setup, which would be used throughout the paper.

When $f(t) \in \mathbb{C}$ is analyzed as $f(t) = \sum_{\omega}A_{\omega}e^{i\omega t}$, $\sum_{\omega}|A_{\omega}| \leq \eta^{\eta}$ and let the required cutoff angular frequency be $1/\eta^{\eta+1}$. Assume that except for zero frequency, no angular frequency $|\omega| < 1/\eta^{\eta+1}$ exists that has non-zero amplitude contributions to $f(t)$. Then conventional filtering required extremely many samples relative to $\eta$, in case we wish to filter away frequencies equal to or greater than cutoff frequency sufficiently such that to nearest integer (for real and imaginary parts) there is zero contribution of these frequencies. 

This is avoided by the following idea. First, lowpass filter away angular frequency $1/2$ to $1$, double frequency of the resulting filter output and then again lowpass filter away angular frequency $1/2$ to $1$. This effectively amounts to filtering out angular frequency $1/4$ to $1/2$. This can be continued such that one filters out angular frequency $1/8$ to $1/4$, $1/16$ to $1/8$ and so forth until $1/\eta^{\eta+1}$ to $2/\eta^{\eta+1}$ is reached.

\subsection{A cycle}
Each filtering process is considered part of a cycle - for example, (effectively) filtering angular frequency $1/2$ to $1$ is part of cycle 1, while $1/4$ to $1/2$ is part of cycle 2, $1/8$ to $1/4$ is cycle 3 and so forth.

However, filters consume samples, so doubling frequency naively would only return us to exponentially many samples relative to $\eta$. Therefore, one needs extrapolation from cycle filter output samples, which then would become next cycle filter input samples.

This idea is inspired from the point that as long as we overcome extrapolation issues that filtered high-frequency parts cause, since high frequency parts are already filtered, one can move onto next frequency ranges.

A cycle $c$, starting from cycle 1, then would first start from filtering cycle input $f_c(t)$ that is lowpass filtered by filter $H_f$. This produces filter output $g_c(t)$, with its samples then used for numerical derivative calculation using differentiation (derivative) filter $H_{d,\mu}$, where $\mu$ refers to derivative order. After numerical derivatives are calculated at $t=T_e$, then one produces an extrapolated value of $g_c(T_e+1)$ via a local polynomial construction, which then in turn are used along with other samples and past extrapolated values to calculate numerical derivatives of $g_c(t)$ at $t=T_e+1$, which then are used to produce an extrapolated value of $g_c(T_e+2)$ and so forth.

\subsection{On extrapolation and numerical derivatives}
The traditional polynomial extrapolation strategy uses a fixed polynomial calculated from existing samples. This leaves no correction mechanism and eventually creates costly extrapolation errors.

A local polynomial extrapolation strategy presented in this paper, by contrast, allows for the correction mechanism by allowing extrapolation errors to be treated as (a combination of) sinusoids. This is done by re-computing numerical derivatives (to order $N_{\mu}$) with a newly extrapolated (single) sample using numerical differentiation filters for each derivative order. This allows treatment of extrapolation errors as having come from a combination of sinusoids, as far as digital filters admit the frequency response interpretation.

\subsection{Why extrapolation-based filtering is possible}
The aforementioned strategies may be questioned for the following reason: given the limited number of pre-extrapolation samples, would not discrete Fourier transform of these samples suggest that valuable information about low frequency signals would be lost?

The answer is that even if we replace $f(t)$ with $f_{DFT}(t)$, where $f_{DFT}(t)$ is constructed from discrete Fourier transform of samples of $f(t)$, the process of doubling frequency for digital frequencies introduces new aliasing such that high frequency parts of $f_{DFT}(t)$ are mingled with low frequency parts of $f_{DFT}(t)$. Therefore, fidelity of $f_{DFT}(t)$ is anyway lost having passed each filtering-extrapolation cycle.

\subsection{Additional terminology}
For each cycle $c$, high-frequencies would refer to post-cutoff frequencies - that is, $|\omega| > \omega_c$. Low-frequencies would refer to pre-cutoff frequencies: $|\omega| \leq \omega_c$. These frequencies are what filter $H_f$ perceive, not `effective' frequencies scaled to original $f(t)$. 

\subsection{On error analysis}
For each cycle, the extrapolation processes would produce deviations from the actual filter $H_f$ output results - this would be called (polynomial) extrapolation errors - though almost. Analysis of polynomial extrapolation errors for a high frequency is a little problematic. First of all, we should really consider entire filtered high frequency signals as errors, since we do not want them. Second, given the initial naive definition for extrapolation errors, heavy errors are reported for high-frequency signals, in contrast to the modified definition where extremely low errors are reported after having adjusted constants appropriately. Therefore, for each cycle, post-cutoff frequency parts of the filter $H_f$ output and their extrapolation errors together are considered extrapolation errors.

The next important point is on the use of Parseval's theorem. Without Parseval's theorem, there is some difficulty in producing tight bounds that connect values of samples with frequency amplitudes - which make frequency analysis difficult. Parseval's theorem therefore is crucial for error analysis.

For extrapolation error analysis, one takes the following strategy. First, at each (extrapolation) sample time $t$, analysis is done as if samples of an ideal function extrapolated are available up to time $t-1$. Then take the initial extrapolation error function produced and consider its extrapolation error and so forth. (`extrapolation error of extrapolation error.') Together with Parseval's theorem, this strategy simplifies extrapolation error analysis significantly.

\subsection{Derivative filter design}
Matching with the goal to minimize high-frequency contributions after filtering and extrapolation, digital differentiation filters $H_{d,\mu}$ are designed as to satisfy frequency response $|H_{d,\mu}(\omega)| < |H_{d,\mu}(\omega_c)|$ for $|\omega| > \omega_c$. The choice of cutoff angular frequency $\omega_c$ would briefly be explained below.

\subsection{On what remains}
The rest of this paper then is about tuning constants (labelled with $k$'s, with different subscripts) so that right balance is to be achieved so that only polynomially many samples are required relative to $\eta$ - for this, see subsection \ref{subsec:filterdesign} and \ref{subsec:constantsk}. The demonstration of the claim depends on three conjectures of Equation \eqref{eqn:filterordernfnd}, Equation \eqref{eqn:filcoeffbound} and Equation \eqref{eqn:ffctimecomplexity} that state the minimal required filter order bound, the filter coefficient magnitude upper bound and the filter coefficient time complexity bound, which are left unproved.

For the choice of the constants, a brief summary may be provided: $k_c=1$ (which by definition gives us `cutoff angular frequency' $\omega_c = 1/\eta$) is determined by Equation \eqref{eqn:gthetabound} when one considers high-frequency and low-frequency extrapolation errors together, $k_{\mu}=2$ (which by definition gives us $N_{\mu}=\eta^2$, the maximum derivative order considered in numerical differentiation) is determined mainly by the low-frequency error equation of Equation \eqref{eqn:lowfqerrdecomp} (considering $|\omega| < \omega_c$), $k_a$ ($1/\eta^{\eta^{k_a}}$ providing minimum magnitude attenuation ratio at $\omega_c$) is determined by Equation \eqref{eqn:wsumintpnp2} at the end of high-frequency and low-frequency extrapolation consolidation analysis. Rest of $k$s then are set by the aforementioned $k$'s so that non-extrapolation errors do not intervene into extrapolation error analysis that were carried out assuming non-extrapolation errors do not exist.

\subsection{Overview of the divide and conquer strategy}
\label{subsec:overviewdivconq}
The divide and conquer strategy is described in Algorithm \ref{alg:divconq}. The general idea goes as follows. At each `cycle' $c$, with input $f_c(t)$ ending with generation of filter output samples that become next cycle input $f_{c+1}(t)$, input $f_c(t)$ is lowpass-filtered by filter $H_f$. (Line 8 to 13 in Algorithm \ref{alg:divconq} - furthermore, $f_1(t) \equiv f(t)$.) This produces filter output $g_c(t)$ from $t=N_f$ to $t=N_f+N_d$ with sample interval $\Delta t = 1$. These samples are used to produce numerical derivative data of $g_c(t)$ at $t=N_f+N_d$, with derivative orders computed to $N_{\mu}$ - that is, numerical computation of $g_c^{(k)}(t)$ from $k=1$ to $k=N_{\mu}$ where $(k)$ refers to numerical derivative order, not exponentiation, at $t=N_f+N_d$ is conducted. This is used to extrapolate $g_c(N_f+N_d+1)$ using a local polynomial constructed from numerical derivatives. This numerically computed $g_c(N_f+N_d+1)$ is then treated as if it is actual $g_c(N_f+N_d+1)$ - this is used again to obtain numerical derivatives of $g_c(t)$ down to order $N_{\mu}$ at $t=N_f+N_d+1$, and then these derivatives are used to construct a local polynomial that is used to obtain extrapolated $g_c(t)$ at $t=N_f+N_d+2$ and so forth. (Line 14 to Line 25) Then we use computed and numerically extrapolated $g_c(t)$ to construct next cycle input $f_{c+1}(t)$ after doubling angular frequency and adjusting the time frame. (Line 27 of Algorithm \ref{alg:divconq})

There are $\eta^2$ cycles - ranging from the first cycle ($c=1$ or $r=1$ in terms of Algorithm \ref{alg:divconq}) to the final $\eta^2$th cycle ($c = \eta^2$ or $r=\eta^2$ in terms of Algorithm \ref{alg:divconq}).

\subsubsection{local polynomial extrapolation}
The idea behind the local extrapolation is this: the usual polynomial interpolation / extrapolation usually accumulates errors rapidly without any taming factors. Why is this so? This is because each polynomial coefficient eseentially encodes derivative information, and derivative is essentially all numerically determined at the single time polynomial interpolation is taken from samples. Because there is no re-evaluation of derivative information, errors simply accumulate.

By contrast, the local extrapolation approach re-evaluates derivative at each sample (discrete) time. Given that the extrapolation interval equals sample interval, where extrapolation interval refers to how far a sample at $t'$ is extrapolated using numerical derivative information at $t$, with $t'-t = \Delta t = 1$, this allows for protection against extrapolation error explosion. Furthermore, extrapolation error in each sample can now be treated as if it comes from an imaginary sinusoid function. This allows numerical derivative filter analysis fully in terms of angular frequency.

We can then think of initial extrapolation error at time $t'$ as extrapolation error arising out of ideal (without numerical errors) samples of $g_c(t)$ from $t=t'-N_d-1$ to $t=t'-1$. And then we treat this initial extrapolation error function as a sinusoid function via discrete Fourier transform from samples ranging from $t=N_f+N_d+1$ to $t=2(N_f+N_d)+N_f-2$. This initial extrapolation error function generates its own extrapolation error function via discrete Fourier transform from error samples ranging from $t=N_f+N_d+2$ to $t=2(N_f+N_d)+N_f-2$ and so forth.

\subsubsection{Derivative filter}
It is well-known that numerical derivative filters have problems with a high-frequency signal, while they can be designed as to achieve high numerical derivative accuracy for a low-frequency signal.

This issue would be evaded by designing numerical derivative filters such that the absolute numerical derivative result for a high-frequency signal is low. This means high numerical derivative inaccuracies, but we do not really care about fidelity of a high-frequency signal, given that we want to filter out high-frequency signals. All we want is to minimize as possible contributions of a high-frequency signal.

\subsubsection{Extrapolation error analysis strategy}
As would be seen, the sub-theme and strategy in extrapolation error analysis is summing up all `extrapolation error of extrapolation error,' along with Parseval's theorem machineries.

\begin{algorithm}
\SetAlgoLined
\SetKwInOut{Input}{input}
\SetKwInOut{Output}{output}
\Input{$f(t)$, coefficient array $h_1$ of digital filter $H_f$ of length $N_f$ indexed from $1$ to $N_f$, coefficient array $h_{2,\mu}$ of derivative filter $H_{d,\mu}$ for derivative order $\mu$ with length of $N_d+1$ indexed from $1$ to $N_d+1$ and $\mu$ ranging from $1$ to $N_{\mu}$, value of $N_f$,$N_d$,$N_{\mu}$}
\Output{$A_0$, zero-frequency amplitude}
 Initialize array $f_c$ of length $N_f+N_d$, index starting from 1\;
 Initialize array $g_c$ of length $2(N_f+N_d)-1$, index starting from $N_f$ to $2(N_f+N_d)+N_f-2$\;
 Initialize array $d$ of length $N_{\mu}$, index starting from $1$\;
 \For{$j \gets 1$ \KwTo $N_f+N_d$}
 {
       $f_c[j] \gets f(j)$\;
 }
 \tcc{For each filtering-extrapolation cycle, with total of $\eta^2$ cycles}
 \For{$r \gets 1$ \KwTo $\eta^2$}
 {
    \tcc{Lowpass filtering}
    \For{$j \gets N_f$ \KwTo $N_f+N_d$}
    {
        $g_c[j] \gets 0$\;
        \For{$m \gets 1$ \KwTo $N_f$}
        {
            $g_c[j] \gets g_c[j] + h_1[m]f_c[j-N_f+m]$\;
        }
    }
    \tcc{Numerical derivative computation down to order $N_{\mu}$ for each time $t$ and extrapolation to $t+1$}
    \For{$j \gets N_f+N_d$ \KwTo $2(N_f+N_d)+N_f-3$}
    {
        \tcc{Numerical derivative computation down to order $N_{\mu}$ for time $t$}
        \For{$\mu \gets 1$ \KwTo $N_{\mu}$}
        {
            $d[\mu] \gets 0$\;
            \For{$m \gets 1$ \KwTo $N_d+1$}
            {
                $d[\mu] \gets d[\mu] + h_{2,\mu}[m]g_c[j-N_d+m-1]$\;
            }
        }
        \tcc{Polynomial extrapolation to $t+1$}
        $g_c[j+1] \gets g_c[j]$\;
        \For{$\mu \gets 1$ \KwTo $N_{\mu}$}
        {
            $g_c[j+1] \gets g_c[j+1] + d[\mu]/\mu!$\;
        }
        
    }
    \tcc{Translating cycle output as new cycle input, after doubling frequency: we need new $N_f+N_d$ samples with $\Delta t=1$ in new frequency/time scale.}
    \For{$j \gets 1$ \KwTo $(N_f+N_d)$}
    {
        $f_c[j] \gets g_c[N_f + 2(j-1)]$\;
    
    }
 
 }
 \tcc{Final computation of $A_0$, zero-frequency amplitude to nearest integer}
 round up $g_c[2(N_f+N_d)+N_f-2]$ to nearest integer: this becomes $A_0$\;

\caption{overview of the divide-and-conquer strategy}
\label{alg:divconq}
\end{algorithm}

\subsection{Number of filtering-extrapolation cycles}
\label{subsec:numbercycle}
First, about the number of cycles. As stated before, the idea is to implement the `divide and conquer' strategy by filtering out $1/2$ of `active' frequencies of original $f(t)$. (That is, we filter out angular frequency $1/2$ to $1$, then $1/4$ to $1/2$, then $1/8$ to $1/4$ and so forth until $1/{\eta}^{\eta+1}$ to $2/{\eta}^{\eta+1}$ is reached.) Since $2^{\eta^2} > \eta^{\eta+1}$, assuming $\eta$ is large enough, $\eta^2$ cycles are sufficient to filter out all meaningful non-zero frequencies of $f(t)$.

This is implemented by doubling frequency of cycle output $p_{c-1}(t)$ of cycle $c-1$ and then re-labelling it as cycle input $f_c(t)$, which is input to filter $H_f$ at cycle $c$.

Some remark about filter output $p_c(t)$. In Algorithm \ref{alg:divconq}, there is no mention about $p_c(t)$. This is because $p_c(t)$ is simply $g_c(t)$ up to $t=N_f+N_d$ (with the implicit assumption that $f_c(t)$ starts from $t=1$ with $\Delta t = 1$), and numerically extrapolated for $t=N_f+N_d+1$ to $t=2(N_d+N_f)+N_f-2$ from available $g_c(t)$. In Algorithm \ref{alg:divconq}, this was simply labelled together as $g_c(t)$ for exposition convenience, but it is better to distinguish actual $g_c(t)$ from numerical version $p_c(t)$.

In the algorithm, $c$ was not used as an index - instead $r$ was used as a cycle index. In the main text, $c$ would indeed be used as cycle index. Thus, $f_1(t)$ refers to the cycle input of the first cycle, $f_2$ refers to the cycle input of the second cycle and so forth. $f_1(t) = f(t)$ by the design of the algorithm.

\subsection{Filter design specifications}
\label{subsec:filterdesign}
We start from specifying filter design parameters, which would aid understanding filter requirements behind Algorithm \ref{alg:divconq}.

For filter $H_f$:
\begin{itemize}
    \item `Cutoff angular frequency' $\omega_c = 1/\eta^{k_c}$: cutoff angular frequency is assumed to have post-cutoff minimum attenuation $a_m$ in filter $H_f$.
    \item Post-cutoff minimum attenuation $a_m = 1/\eta^{\eta^{k_a}}$: A signal of angular frequency $|\omega| \geq \omega_c$ is attenuated by at least $a_m$ of original amplitude by filter $H_f$. For example, if the original signal is $e^{i\omega t}$ with $|\omega| \geq \omega_c$, then $H_f$ filter output is, in maximum magnitude, less than or equal to $a_m$.
    \item The requirement that lowpass filter has no overshoot in the lowpass region. That is, frequency response $|H(\omega)| < 1$ for $|\omega| \neq 0$. Furthermore, $H(\omega) = 1$ for $\omega = 0$.
\end{itemize}
For derivative filters $H_{d,\mu}$:
\begin{itemize}
    \item Subscript $\mu$ refers to derivative order. For example, taking numerical differentiation of $f(t)$ of derivative order $\mu=2$ means numerically calculating $f''(t)$. 
    \item `Cutoff angular frequency' $\omega_c = 1/\eta^{k_c}$: $\omega_c$ is shared with filter $H_f$. Meaning of cutoff angular frequency, however, changes. For derivative filters, up to angular frequency $\omega_c$, derivative of $e^{i\omega t}$ is very accurate up to maximum error magnitude $e_d$. For $|\omega| > \omega_c$, frequency response $|H_{d,\mu}(\omega)| < |H_{d,\mu}(\omega_c)|$.
    \item Maximum error magnitude when obtaining numerical derivative of $e^{i\omega t}$ for $|\omega| \leq \omega_c$: $e_d = 1/\eta^{\eta^{k_d}}$: deviation magnitude from actual numerical derivative (of any derivative order $\mu$) is bounded above by $e_d$.
\end{itemize}
It would be assumed, for further convenience, that $k_a = k_d$.

The reason behind particular form of $\omega_c$ would be explained when extrapolation error issues are discussed, along with choices for $k$ constants. The required filter order(s) for $H_f$ and $H_{d,\mu}$ are then stated as:
\begin{equation}
\label{eqn:filterordernfnd}
N_f,N_d \leq \eta^{k_a k_{aa} + k_c k_{ca}}
\end{equation}
where $k_{aa}$ and $k_{ca}$ are other constants. In this paper, this bound would be unproved, left to the subsequent paper, along with the question of proving polynomial time complexity in obtaining filter coefficients.

It is further assumed that filter coefficients are bounded as:
\begin{equation}
\label{eqn:filcoeffbound}
|h[i]| \leq 2^{\eta^{k_a k_{ab} + k_c k_{cb}}}
\end{equation}
where $h[i]$ refers to any filter coefficient $h_{2,\mu}[m]$ or $h_1[m]$ in Algorithm \ref{alg:divconq}. The maximum number $K_{id}$ of integer (non-fractional) digits possible for any number variable used in Algorithm \ref{alg:algxt} and \ref{alg:divconq} then is, via Line 11 and Line 18 of Algorithm \ref{alg:divconq}:

\begin{equation}
\label{eqn:kid}
K_{id} \leq 4\eta^{k_a k_{ab} + k_c k_{cb}}
\end{equation}
Then,
\begin{equation}
\label{eqn:kbaikbmi}
k_{bai} = k_{bmi} = 2(k_a k_{ab} + k_c k_{cb})
\end{equation}
$\eta^{k_{bai}}$ would refer to number of integer binary digits addition computation would keep. $\eta^{k_{bmi}}$ would refer to number of integer binary digits multiplication computation would keep.

Time complexity of obtaining each filter coefficient is assumed to be bounded as:
\begin{equation}
\label{eqn:ffctimecomplexity}
O(\eta^{k_a k_{ac} + k_c k_{cc} + k_p k_{pc}})
\end{equation}
where $1/\eta^{\eta^{k_p}}$ refers to the maximum filter coefficient error magnitude tolerated. $k_{ffc}$ is then defined as:
\begin{equation}
\label{eqn:kffc}
k_{ffc} \equiv k_a k_{ac} + k_c k_{cc} + k_p k_{pc}
\end{equation}
Note that $k_{aa}$, $k_{ca}$, $k_{ab}$, $k_{cb}$, $k_{ac}$, $k_{cc}$ and $k_{pc}$ are all undetermined constants.

\subsection{General error analysis}
\label{subsec:err}
\subsubsection{What is error?}
\label{subsubsec:errdef}
We need to define what would be considered error when considering extrapolation or filter outputs. Things are initially straightforward. One first can consider various sources of errors - addition computation errors, multiplication computation errors, filter coefficient computation errors and extrapolation errors. They all have clear ideal references. 

However, our aim is to compute $A_0 \equiv n_h$, zero-frequency amplitude to nearest integer and the number of Hamiltonian paths. Therefore, filtered high frequency signals, demarcated by cutoff frequency $\omega_c$, should be considered errors, since our aim is that all angular frequencies except zero are filtered out.

\subsubsection{General error analysis, continued}
\label{subsubsec:conterroranalysis}
For any type of errors induced by numerical approaches, one can think of them as generated by sinusoids. Why this is possible is given by discrete Fourier transform analysis.

\begin{itemize}
\item First, there are filter $H_f$ output errors, which are caused by a combination of multiplication, addition computation errors and filter coefficient computation errors. Line 11, 18 and 23 in Algorithm \ref{alg:divconq} are the sources of the errors.
\item The next type of error is filter $H_{d,\mu}$ errors in computing numerical derivatives of order $\mu$. But error analysis for numerical derivative of a high-frequency ($|\omega|>\omega_c$) signal would have to be thought differently - it is not divergence from actual derivative that matters but computed numerical derivatives that matter. We intend to make high-frequency numerical derivatives as small as possible without sacrificing faithful polynomial extrapolation capacities. Numerical derivative errors themselves do not matter until polynomial extrapolation is conducted using these numerical derivatives.
\item Basically, the point is that we want none of high-frequency components - thus, high-frequency components are basically considered as errors.
\item Numerical derivative computation errors only become relevant when polynomial extrapolation is carried out in Line 23 of Algorithm \ref{alg:divconq}. (Note that numerical differentiation is carried out in Line 18 of Algorithm \ref{alg:divconq}.)
\item Even when there is no numerical differentiation error, the fact that we use polynomials to extrapolate means there would be inherent extrapolation error given by polynomial degree $N_{\mu}$, relevant for Line 23 of Algorithm \ref{alg:divconq}.
\item Initial errors in $f_c(t)$ for $c \geq 2$. At each cycle, errors accumulated from the previous cycle $c-1$ become the errors in $f_c(t)$. They would be treated as a new sinusoidal input contribution in error analysis. 
\item Then there is error in computing $f(t) \equiv f_1(t)$. Line 4 and 14 in Algorithm \ref{alg:algxt} contain the source of sinusoidal ($e^{i\omega t}$) calculation errors. Line 11 and 21 contain the source of addition errors. Line 14 contains the source of multiplication errors.
\end{itemize}
It would initially be assumed that the sole type of errors is polynomial extrapolation error, which is induced by numerical derivative error that becomes polynomial coefficient error and also induced by inherent finite-degree polynomial extrapolation error. Remaining types of errors are revisited later. Note that filter output errors become next cycle input errors.

\subsubsection{Error analysis strategy}
\label{subsubsec:errstrategy}
\begin{itemize}
\item The key functions to be used are $\theta_c(t)$, $g'_{\theta_c}(t)$, $g_{\theta_c}(t)$, $w_{\theta_c,j}(t)$ and $w'_{\theta_c,j}(t)$. $c$ subscript refers to the fact that error analysis is done per one cycle.
\item $\theta_c(t)$ stands for cycle input - or filter input to filter $H_f$ at cycle $c$.
\item $g'_{\theta_c}(t)$ is function to be extrapolated, originated as filter output when filter input is $\theta_c(t)$ under filter $H_f$. 
\item Let $g'_{\theta_c}(t) = \gamma_{\theta_c}(t) + \psi'_{\theta_c}(t)$, where $\gamma_{\theta_c}(t)$ is the low frequency part(s) of $g'_{\theta_c}(t)$ and $\psi'_{\theta_c}(t)$ is the high frequency part(s) of $g'_{\theta_c}(t)$, demarcated by angular frequency $\omega_c$. Note that angular frequency decomposition of $g'_{\theta_c}(t)$ comes from full continuous Fourier transform from $t=-\infty$ to $t=\infty$.
\item Let $g_{\theta_c}(t) = \gamma_{\theta_c}(t) + \psi_{\theta_c}(t)$.
\item $\psi_{\theta_c}(t) = \psi'_{\theta_c}(t)$ for positive integer $t < z-1$ and $\psi_{\theta_c}(t) = \psi'_{\theta_c}(z-1)$ for positive integer $t \geq z-1$, where $z = N_f+N_d+1$. While $\psi_{\theta_c}(t)$ is incompletely defined, this is not a problem, since we would treat $\psi_{\theta_c}(t)$ as if it is a periodic signal of period $z_2$, given by discrete Fourier transform from samples ranging from $t=1$ to $t=z_2$.
\item Change from $g'_{\theta_c}(t)$ to $g_{\theta_c}(t)$ reflects the view that high frequency signals are to be treated as errors. Furthermore, since samples of any function are only available from $t=1$ to $t=N_f+N_d$, there is freedom as to what function really is being extrapolated. For low frequency signals, this freedom cannot be exploited, since we need fidelity of these signals to extract $A_0$, the zero frequency amplitude of $f(t)$. By contrast, high frequency signals are filtered ones and we do not need them. Essence is that extrapolating a function at time $t$ starts from the value of the function at time $t-1$ (this is why $\psi_{\theta_c}(t) = \psi'_{\theta_c}(z-1)$), and we add to this value using numerical derivatives computed at time $t-1$, which then becomes the extrapolation result. This point would become clear when defining $\psi_{\theta_c,j}(t)$ and especially $\upsilon_{\theta_c,j}(t)$, which roughly acts as the additional and numerical extrapolation contribution to $\psi'_{\theta_c,j-1}(j-1)$ when extrapolating $\psi'_{\theta_c,j-1}(t)$ at $t=j$, with $j \geq z+1$. The full definition of $\upsilon_{\theta_c,j}(t)$ is given by Equation \eqref{eqn:upsilondefff}, which roughly says that $\upsilon_{\theta_c,j}(t)$ is additional extrapolation contribution to $\psi_{\theta_c,j-1}(t-1)$ for positive integer $t \geq j$, assuming samples of $\psi_{\theta_c,j-1}(t)$ are available from time $1$ to $t-1$ without inaccuracies. Given that $\upsilon_{\theta_c,j}(t)$ itself has to be `extrapolated', one does not have to worry so much about $t > j$, except to note that this is a convenient trick for error analysis.
\item Let us think of initial extrapolation error at time $t'$ when extrapolating $g_{\theta_c}(t)$, assuming samples of $g_{\theta_c}(t)$ from $t=1$ to $t=t'-1$ are available without inaccuracies. This allows us to formulate function $w'_{\theta_c,z}(t)$, where $z=N_f+N_d+1$ refers to the starting point of non-zero extrapolation errors.

However, the high-frequency part of $g_{\theta_c}(t)$ produces large extrapolation errors, despite actual extrapolation results not being so. Therefore, it is better to think of the high-frequency part itself as if it is error.

This means forming $w'_{\theta_c,z}(t)$ from $g_{\theta_c}(t)$ as, with $z$ defined as $z = N_f+N_d+1$:
\begin{equation}
\label{eqn:wthetainitial}
w'_{\theta_c,z}(t) = \zeta_{\theta_c,z}(t) + \upsilon_{\theta_c,z}(t) = \gamma_{\theta_c,z}(t) + \psi'_{\theta_c,z}(t)
\end{equation}
where $\zeta_{\theta_c,z}(t)$ refers to extrapolation error for $\gamma_{\theta_c}(t)$ assuming that samples of $\gamma_{\theta_c}(t)$ are available from time $1$ to $t-1$ without inaccuracies, $\upsilon_{\theta_c,z}(t)$ contains $\psi_{\theta_c}(t)$ and additional extrapolation output when extrapolating $\psi_{\theta_c}(t)$, assuming samples of $\psi_{\theta_c}(t)$ are available from time $1$ to $t-1$ without inaccuracies. $\gamma_{\theta_c,z}(t)$ is the low-frequency part(s) of $w'_{\theta_c,z}(t)$ and $\psi'_{\theta_c,z}(t)$ is the high frequency part(s) of $w'_{\theta_c,z}(t)$, demarcated by angular frequency $\omega_c$. Note that Fourier decomposition is given by discrete Fourier transform using samples from $t=1$ to $t=z_2$.

If $\gamma_{\theta_c}(t)$ is given by $\sum_{\omega}A_{\omega}e^{i\omega t}$, then for positive integer $t \geq z$ (otherwise for positive integer $t<z$, $\zeta_{\theta_c,z}(t) = 0$):
\begin{multline}
\label{eqn:lowfqadderr}
\zeta_{\theta_c,z}(t) = \sum_{\omega}A_{\omega}e^{i\omega (t-1)}\\\left(\sum_{k=1}^{N_{\mu}}\frac{e_{d,k,\omega}}{k!} + (i\omega)^{N_{\mu} + 1}e^{i\omega t_{\xi}}\frac{1}{(N_{\mu} + 1)!} \right)
\end{multline}
where $e_{d,k,\omega}$ refers to error in computing numerical derivative of order $k$ for $e^{i\omega t}$ at $t=0$, and $0 \leq t_{\xi} \leq 1$. $A_{\omega}e^{i\omega (t-1)}\sum_{k=1}^{N_{\mu}}\frac{e_{d,k,\omega}}{k!}$ refers to the errors induced by errors in polynomial coefficients when extrapolating $A_{\omega}e^{i\omega t}$, which in turn are about errors in computing numerical derivative. $A_{\omega} e^{i\omega (t-1)}(i\omega)^{N_{\mu} + 1}e^{i\omega t_{\xi}}\frac{1}{(N_{\mu} + 1)!}$ refers to inherent polynomial extrapolation error, given by Taylor's theorem.

If $\psi_{\theta_c}(t)$ is given by $\sum_{\omega}A_{\omega}e^{i\omega t}$, then for positive integer $t \geq z$ (otherwise for positive integer $t<z$, $\upsilon_{\theta_c,z}(t) = 0$):
\begin{equation}
\label{eqn:highfqadderr}
\upsilon_{\theta_c,z}(t) = \sum_{\omega}A_{\omega}e^{i\omega (t-1)} + \sum_{\omega} A_{\omega}e^{i\omega(t-1)}\sum_{k=1}^{N_{\mu}}d_{n,k,\omega}\frac{1}{k!}
\end{equation}
where $d_{n,k,\omega}$ is numerical derivative of $e^{i\omega t}$ of order $k$ at $t=0$.

Let $\psi_{\theta_c,z}(t)$ be formed from $\psi'_{\theta_c,z}(t)$ by:
\begin{equation}
\label{eqn:wwapostro1}
\psi_{\theta_c,z}(t) = \psi'_{\theta_c,z}(t)
\end{equation}
for positive integer $t < z$ and
\begin{equation}
\label{eqn:wwapostro2}
\psi_{\theta_c,z}(t) = \psi'_{\theta_c,z}(z)
\end{equation}
for positive integer $t \geq z$. This change from $\psi'_{\theta_c,z}$ to $\psi_{\theta_c,z}$ does not pose any issue in that given non-existence of samples of any function from $t=z$, we can choose arbitrary values as the `right' result for $t=z$ to $t=2(N_f+N_d)+N_f-2$.

Furthermore, form of Equation \eqref{eqn:wwapostro2} helps in that it suggests we are really counting down additional contributions to the extrapolation result for high frequency signals when moving from $t-1$ to $t$, not usual extrapolation errors. That is, $\psi_{\theta_c,z}$ really just keeps (or maintains) the initial extrapolation contribution at time $z$, which is simply $\psi'_{\theta_c,z}(z)$ for positive integer $t \geq z$. In short, aforementioned functional form allows invoking Equation \eqref{eqn:upsilontheta} when conducting `high frequency' error analysis calculations. Let us now define $w_{\theta_c,z}(t)$:
\begin{equation}
\label{eqn:wunprimedz}
w_{\theta_c,z}(t) = \gamma_{\theta_c,z}(t) + \psi_{\theta_c,z}(t)
\end{equation}

Then $\psi_{\theta_c,j}(t)$ for $j>z$ would be defined in general sense: $\psi_{\theta_c,j}(t)$ maintains the `additional' extrapolation contribution at time $j$ for positive integer $j > z$. That is,
\begin{equation}
\label{eqn:wwwvv1}
\psi_{\theta_c,j}(t) = \psi'_{\theta_c,j}(j)
\end{equation}
for positive integer $t \geq j$. For other positive integer $t < j$, $\psi_{\theta_c,j}(t) = \psi'_{\theta_c,j}(t)$.

But $\psi'_{\theta_c,j}(t)$ for $j>z$ has not been defined yet - this would be provided later. Let $w'_{\theta_c,j}(t)$ for positive integer $j>z$ satisfy:
\begin{equation}
\label{eqn:wwwvvgen}
w'_{\theta_c,j}(t) = \zeta_{\theta_c,j}(t) + \upsilon_{\theta_c,j}(t) = \gamma_{\theta_c,j}(t) + \psi'_{\theta_c,j}(t)
\end{equation}
where $\zeta_{\theta_c,j}(t)$ is extrapolation error for extrapolating $\gamma_{\theta_c,j-1}(t)$ (which would be defined in the below), assuming samples of $\gamma_{\theta_c,j-1}$ are available without inaccuracies from time $1$ to time $t-1$. ($\zeta_{\theta_c,j}(t) = 0$ for positive integer $t <j$.) $\gamma_{\theta_c,j}(t)$ is defined as the low frequency components of $w'_{\theta_c,j}(t)$, demarcated by $|\omega| \leq \omega_c$, with $\psi'_{\theta_c,j}(t)$ defined as the high frequency components of $w'_{\theta_c,j}(t)$.

$\upsilon_{\theta_c,j}(t)$ is defined for positive integer $t \geq j$ (and $j > z$) as:
\begin{equation}
\label{eqn:upsilondefff}
\upsilon_{\theta_c,j}(t) = \psi_{n,\theta_c,j-1}(t) - \psi_{\theta_c,j-1}(t-1)
\end{equation}
where $\psi_{n,\theta_c,j-1}(t)$ refers to the extrapolation result (not error) of extrapolating $\psi_{\theta_c,j-1}(t)$ using samples of $\psi_{\theta_c,j-1}$ from time $1$ to time $t-1$ that are assumed to be available without inaccuracies. 

Otherwise for positive integer $t$,
\begin{equation}
\label{eqn:upsilondefff3}
\upsilon_{\theta_c,j}(t) = 0
\end{equation}
This completes defining $\upsilon_{\theta_c,j}(t)$ as the additional extrapolation contribution induced at time $j$ when extrapolating $\psi_{\theta_c,j-1}(t)$.

Now let us refer back to Equation \eqref{eqn:wwwvvgen}. $\gamma_{\theta_c,j}(t)$ is defined as the low frequency part of $w'_{\theta_c,j}(t)$, along with $\psi'_{\theta_c,j}(t)$ defined as the high frequency part of $w'_{\theta_c,j}(t)$, demarcated by $|\omega| > \omega_c$ (for low frequency, $|\omega| \leq \omega_c$). Then, we define $w_{\theta_c,j}(t)$ for $j > z$ by:
\begin{equation}
\label{eqn:wgengengen}
w_{\theta_c,j}(t) = \gamma_{\theta_c,j}(t) + \psi_{\theta_c,j}(t)
\end{equation}
with $\psi_{\theta_c,j}(t)$ satisfying:
\begin{equation}
\label{eqn:psigengen}
\psi_{\theta_c,j}(t) = \psi'_{\theta_c,j}(j)
\end{equation}
for positive integer $t \geq j$ and
\begin{equation}
\label{eqn:psigengen2}
\psi_{\theta_c,j}(t) = \psi'_{\theta_c,j}(t)
\end{equation}
for positive integer $t < j$.

This completes the setup required for error analysis calculations. 

\item The total extrapolation `error', at the end of cycle $c$, provided filter input $\theta_c(t)$ at cycle $c$, accounting for the high frequency modification, is then given by $\sum_{j=z}^{z_2}w_{\theta_c,j}(t)$. 
\end{itemize}

\subsubsection{`Low-frequency' extrapolation error}
\label{subsubsec:lowfreqexterr}
Let us examine $w_{\theta_c,m}(t)$ for positive integer $m \geq z$. Given the discrete Fourier transform decomposition such that
\begin{equation}
\label{eqn:lowfqgamma}
\gamma_{\theta_c,m}(t) = \sum_{\omega} A_{\omega}e^{i\omega t}
\end{equation}
the following holds for positive integer $t \geq m+1$:
\begin{multline}
\label{eqn:lowfqerrdecomp}
\zeta_{\theta_c,m+1}(t) = \sum_{\omega}A_{\omega}e^{i\omega (t-1) }\\\left(\sum_{k=1}^{N_{\mu}}\frac{e_{d,k,\omega}}{k!} + (i\omega)^{N_{\mu} + 1}e^{i\omega t_{\xi}}\frac{1}{(N_{\mu} + 1)!} \right)
\end{multline}
where $e_{d,k,\omega}$ refers to error in computing numerical derivative of order $k$ for $e^{i\omega t}$ at $t=0$, and $0 \leq t_{\xi} \leq 1$. Let us examine Equation \eqref{eqn:lowfqerrdecomp}. $A_{\omega}e^{i\omega (t-1)}\sum_{k=1}^{N_{\mu}}\frac{e_{d,k,\omega}}{k!}$ refers to the errors induced by errors in polynomial coefficients when extrapolating $A_{\omega}e^{i\omega t}$, which in turn are about errors in computing numerical derivative. $A_{\omega}e^{i\omega (t-1)}(i\omega)^{N_{\mu} + 1}e^{i\omega t_{\xi}}\frac{1}{(N_{\mu} + 1)!}$ refers to inherent polynomial extrapolation error, given by Taylor's theorem. We now see that as far as $e_{d,k,\omega}$ is small enough and $N_{\mu}$ is large enough, the low-frequency extrapolation error can be ignored - as far as the high-frequency contribution of $\zeta_{\theta_c,m+1}(t)$ to $w_{\theta_c,m+2}(t)$ is small enough. This would be confirmed in the following discussion.

As aforementioned, let $e_{d,k,\omega}$ be bounded as, applying only for $|\omega| \leq \omega_c$:
\begin{equation}
\label{eqn:djeomegabound}
|e_{d,k,\omega}| \leq \frac{1}{\eta^{\eta^{k_{d}}}}
\end{equation}
where $e_d = \frac{1}{\eta^{\eta^{k_{d}}}}$ as a convenient shortcut.

\subsubsection{`High-frequency' extrapolation error}
\label{subsubsec:highfreqexterr}
Given high-frequency signal $\psi'_{\theta_c,m}(t)$, $\psi_{\theta_{c.m}}(t)$ is analyzed with discrete Fourier transform, transform-relevant samples ranging from $t=1$ to $t=z_2$, with $m \geq z$: 
\begin{equation}
\label{eqn:psitheta}
\psi_{\theta_c,m}(t) = \sum_{\omega} A_{\omega}e^{i\omega t}
\end{equation}
Then the following holds for positive integer $t \geq m+1$:
\begin{equation}
\label{eqn:upsilontheta}
\upsilon_{\theta_c,m+1}(t) = \sum_{\omega} A_{\omega}e^{i\omega (t-1)}\sum_{k=1}^{N_{\mu}}d_{n,k,\omega}\frac{1}{k!}
\end{equation}
where $d_{n,k,\omega}$ is numerical derivative of $e^{i\omega t}$ of order $k$ at $t=0$. $d_{n,k,\omega}$ is bounded by:
\begin{equation}
\label{eqn:highfqndbound}
|d_{n,k,\omega}| \leq {\omega_c}^{k}
\end{equation}
which follows from the derivative filter requirement that frequency response be $|H(\omega)| < |H(\omega_c)|$ for $|\omega| \neq \omega_c$.

We intend to minimize magnitude of $\upsilon_{\theta_c,m+1}(t)$. Or say, we want to make $|\upsilon_{\theta_c,m+1}(t)| \ll 1$.

\subsubsection{Parseval's theorem}
\label{subsubsec:parseval}
We would invoke Parseval's theorem heavily in following error analysis. Parseval's theorem states that:
\begin{equation}
\label{eqn:parseval}
\frac{1}{N}\sum_{t=1}^N|u(t)|^2 = \sum_{\omega}|U(\omega)|^2
\end{equation}
where $N$ represents the total number of samples assumed to be available.

\subsubsection{Consolidating both `high-frequency' and `low-frequency' errors}
\label{subsubsec:consolidate}
Let us consolidate both high-frequency and low-frequency extrapolation errors. From how $w_{\theta_c,j}(t)$ is defined , the following is the magnitude upper bound on $w_{\theta_c,j+1}(t)$ for sufficiently large $\eta$:
\begin{equation}
\label{eqn:wnewbound1}
\max\left[\sum_{t=1}^{z_2}|w_{\theta_c,j+1}(t)|^2\right] < (2\omega_c)^2\max\left[\sum_{t=1}^{z_2}|w_{\theta_c,j}(t)|^2\right]
\end{equation}
where $z_2 = 2(N_f+N_d)+N_f-2$. Equation \eqref{eqn:wnewbound1} comes from the fact that $d_{n,k,\omega}$ in Equation \eqref{eqn:upsilontheta} is bounded by Equation \eqref{eqn:highfqndbound}. 

Therefore, we make the following observation. Let $\theta_c(t)$ be decomposed via Fourier transform as:
\begin{equation}
\label{eqn:thetactdiv}
\theta_c(t) = \sum_{\omega}A_{\omega}e^{i\omega t}
\end{equation}
with $\sum_{\omega}|A_{\omega}| \leq \eta^{\eta}$. Then,
\begin{multline}
\label{eqn:gthetabound}
\frac{1}{z_2}\sum_{t=1}^{z_2}|w_{\theta_c,z}(t)|^2 \leq \\ \sum_{\omega}|A_{\omega}|^2\left(\frac{3+2\omega_c}{\eta^{\eta^{k_a}}}\right)^2 < \sum_{\omega}|A_{\omega}|^2\left(\frac{4}{\eta^{\eta^{k_a}}}\right)^2
\end{multline}
where $z = N_f+N_d+1$ and $z_2 = 2(N_f+N_d)+N_f-2$ with Parseval's theorem implicitly invoked. First, $A_{\omega}e^{i\omega t}$ in $\theta_c(t)$ for angular frequency $\omega > \omega_c$ decays in magnitude to $|A_{\omega}|/\eta^{\eta^{k_a}}$ in $g_{\theta_c}(t)$. The contribution to $w_{\theta_c,m}$ induced by this high-frequency ($|\omega| > \omega_c$) part of $g_{\theta_c}$ is the high-frequency part plus the additional error contribution given by Equation \eqref{eqn:upsilontheta}. This gives the upper bound in magnitude of the total contribution as $(1+2\omega_c)|A_{\omega}|/\eta^{\eta^{k_a}}$. Furthermore, for $A_{\omega}e^{i\omega t}$ in $\theta_c(t)$ for angular frequency $\omega \leq \omega_c$, Equation \eqref{eqn:djeomegabound} along with Equation \eqref{eqn:lowfqerrdecomp} suggests that its contribution to $w_{\theta_c,z}$ is upper bounded in magnitude by $2|A_{\omega}|/\eta^{\eta^{k_a}}$, given that $k_a=k_d$, as assumed.

As far as $\omega_c \ll 1$ (by which it would mean $\omega_c = 1/\eta$ with the default assumption of large $\eta$), as guaranteed, the following upper bound holds from Equation \eqref{eqn:wnewbound1}:
\begin{equation}
\label{eqn:wsumintpnp1}
\max\left[\sum_{m=m'+1}^{z_2}\sum_{t=1}^{z_2}|w_{\theta_c,m}(t)|^2\right] \leq  \max\left[\frac{\sum_{t=1}^{z_2}|w_{\theta_c,m'}(t)|^2}{1-(2\omega_c)^2}\right]
\end{equation}

\begin{equation}
\label{eqn:wsumintpnp2}
\frac{1}{z_2}\sum_{m=z}^{z_2}\sum_{t=1}^{z_2}|w_{\theta_c,m}(t)|^2 < \left(\frac{5}{\eta^{\eta^{k_a}}}\right)^2\sum_{\omega}|A_{\omega}|^2
\end{equation}

Equation \eqref{eqn:wsumintpnp2} provided magnitude upper bound on extrapolation error contributions to the next cycle input, having assumed that there was no error contribution to the present cycle input.

This establishes that as far as $k_{a} \gg 1$ (by which it would mean $k_a =2$) and $k_{\mu} \gg 1$ (by which it would mean $k_{\mu} = 2$ and where $\eta^{k_{\mu}} \equiv N_{\mu}$), extrapolation error does not pose an issue for Algorithm \ref{alg:divconq}.

\subsubsection{Settings for analysis of other types of errors}
\label{subsubsec:setothertypeerror}
Having considered per-cycle extrapolation errors, we now consider other types of errors - sinusoid computation errors, addition computation errors, multiplication computation errors and filter coefficient errors.

General settings would be discussed below. First, we need to define how many integer and fractional bits we are going to maintain for addition and multiplication. 
\begin{itemize}
\item The requirement of maintaining $\eta^{k_{bai}}$ integer bits (for each of real and complex parts) and $\eta^{k_{baf}}$ fractional bits for each sum (addition) of two numbers is imposed. This assumes that each number involved has maximum of $\eta^{k_{bai}} - 1$ integer bits.
\item The requirement of maintaining $\eta^{k_{bmi}}$ integer digits and $\eta^{k_{bmf}}$ fractional digits per each product (multiplication) of two numbers is imposed. This assumes that each number involved has maximum of $(\eta^{k_{bmi}}-2)/2$ integer digits.
\item To simplify analysis, $k_{bai} = k_{bmi}$. $k_{baf} +2 = k_{bmf}$ would be imposed for a reason explained later.
\end{itemize}
The above requirement simplifies error analysis in sense that we may separate addition computation error and multiplication computation error analysis from sinusoidal calculation error analysis for Algorithm \ref{alg:algxt} - to be discussed in subsubsection \ref{subsubsec:xtsinuerr}. Eventually, the aim is to only leave addition errors and extrapolation errors as of only concerns for Algorithm \ref{alg:algxt} and \ref{alg:divconq}.

\subsubsection{Errors in computing x(t) and f(t): sinusoid computation error}
\label{subsubsec:xtsinuerr}
Line 4 and 14 in Algorithm \ref{alg:algxt} contain sources of sinusoid computation errors. 

Line 4 sinusoid computation errors can be brushed aside, given Line 11 (and Line 21, in that they are considered together) in the same algorithm, when:
\begin{equation}
\label{eqn:sinvsadd}
\frac{1}{\eta^{\eta^{k_s}}} \ll \frac{2}{2^{\eta^{k_{baf}}}}
\end{equation}
where the left-hand side refers to the maximum error magnitude when computing sinusoid $e^{i\omega t}$.

Line 14 sinusoid computation errors can be brushed aside, as long as:
\begin{equation}
\label{eqn:sinvsmult}
\frac{\max |u_2[m]|}{\eta^{\eta^{k_s}}} \ll \frac{2}{2^{\eta^{k_{bmf}}}}
\end{equation}
where $\max |u_2[m]|$ is assumed to be $\eta^{\eta +1}$, accounting for errors. In the ideal case without other errors, $\max |u_2[m]| \leq \eta^{\eta}$.

\subsubsection{Errors in computing x(t) and f(t): multiplication error}
\label{subsubsec:ftmulterr}
Multiplication errors, for Algorithm \ref{alg:algxt}, occur in Line 14. We may simplify error analysis for the multiplication error contribution to $x(t)$ (and $f(t)$) as:
\begin{equation}
\label{eqn:multerrorxt}
|e_{mx}(t)| \leq \eta \left(\eta^{\eta}\right)\frac{2}{2^{\eta^{k_{bmf}}}}
\end{equation}
The first $\eta$ in the right-hand side of Equation \eqref{eqn:multerrorxt} comes from the number of levels in the algorithm - referring to the number of levels that new multiplication errors may be produced. $2/2^{\eta^{k_{bmf}}}$ refers to maximum error magnitude induced per each multiplication. Additional $\eta^{\eta}$ refers to how the initial multiplication error in Line 14 creates `explosively' accumulates as it moves through the algorithm \ref{alg:algxt}. The initial errors are summed up at maximum of $\eta$ times per each vertex per one level. Then the result is summed up at maximum of $\eta$ times in the next level, and there are $\eta$ levels. Thus, $\eta^{\eta}$.

\subsubsection{Errors in computing x(t) and f(t): addition error}
\label{subsubsec:xtadderr}
\begin{equation}
\label{eqn:adderrorxt}
|e_{ax}(t)| \leq \eta \left(\eta^{2}\right)^{\eta}\frac{2}{2^{\eta^{k_{baf}}}}
\end{equation}
where $e_{ax}$ refers to the addition error contribution to $x(t)$ (and in turn, $f(t)$). First $\eta$ again refers to the number of levels. $2/2^{\eta^{k_{baf}}}$ refers to maximum error magnitude induced per each addition. $\eta^2$ refers to the fact that Line 11 in Algorithm \ref{alg:algxt} has two inner loops with index from $1$ to $\eta$. The exponential $\eta$ refers to the number of levels that errors exponentially build up, with Line 11 and 21 considered together. 

\subsubsection{Errors in per-cycle computation: filter coefficient error}
\label{subsubsec:divconqfcerr}
Let the cycle input (in Algorithm \ref{alg:divconq}) be $\theta_c(t)$ (that is, assume as if $f_c(t) = \theta_c(t)$). Furthermore, assume $|\theta_c(t)| \leq \eta^{\eta}$. The goal here is to make an filter coefficient error at its occurrence less than multiplication and addition errors in magnitude as to allow us to ignore filter coefficient errors.

Line 11 of Algorithm \ref{alg:divconq} has the first occurrence of filter coefficient errors. In order to ignore the filter coefficient error and consider only the multiplication error in the line,
\begin{equation}
\label{eqn:fcerrmulerr}
\frac{1}{2^{\eta^{k_p}}} |\theta_{cm}| < \frac{2}{2^{\eta^{k_{bmf}}}}
\end{equation}
where $\theta_{cm}$ represents the maximum magnitude of $\theta_c(t)$, which we may count as $\eta^{\eta}$. $1/2^{\eta^{k_p}}$ in the left-hand side of Equation \eqref{eqn:fcerrmulerr} represents maximum error magnitude in computing each filter coefficient. 

Analysis for Line 18 of Algorithm \ref{alg:divconq} remains same as analysis for Line 11.

\subsubsection{Errors in per-cycle computation: multiplication error}
\label{subsubsec:divconqmulterr}
The issue to be discussed here is how to make multiplication error magnitude less than addition error magnitude.

Line 11 of Algorithm \ref{alg:divconq} contains the first occurrence of multiplication errors. To make the addition error cover up the multiplication error in the line,
\begin{equation}
\label{eqn:mulerradderr1}
\frac{2}{2^{\eta^{k_{bmf}}}} < \frac{1}{2^{\eta^{k_{baf}}}}
\end{equation}
Line 18 and Line 23 share same aforementioned analysis.

\subsubsection{Errors in per-cycle computation: addition errors without extrapolation errors}
\label{subsubsec:divconqaddextrerr}
For subsubsection \ref{subsubsec:line11adderr}, \ref{subsubsec:line18adderr} and \ref{subsubsec:line23adderr} discussing addition errors in Algorithm \ref{alg:divconq}, it would initially be assumed that there is no extrapolation error. Extrapolation errors induced by addition errors would be discussed separately.

\subsubsection{Errors in per-cycle computation: Line 11 addition error}
\label{subsubsec:line11adderr}
Line 11 of Algorithm \ref{alg:divconq} accumulates, per each $g_{\theta_c}(t)$ (or as if $g_c(t)$ in the algorithm satisfies $g_c(t) = g_{\theta_c}(t)$), the following addition error maximum magnitude at the end of Line 13:
\begin{equation}
\label{eqn:addaccum1}
\frac{2N_f}{2^{\eta^{k_{baf}}}}
\end{equation}

\subsubsection{Errors in per-cycle computation: Line 18 addition error}
\label{subsubsec:line18adderr}
Line 18 in Algorithm \ref{alg:divconq} accumulates the following maximum addition error magnitude at the end of Line 20 for each $d[\mu]$:
\begin{equation}
\label{eqn:addaccum3}
\frac{2(N_d+1)}{2^{\eta^{k_{baf}}}}
\end{equation}
This contributes to $g_c[j+1]$ at Line 23 (though not looping for all possible $j$ - one looks at how $d[\mu]$ at each $j$ affects $g_c[j+1]$) with the following maximum magnitude:
\begin{equation}
\label{eqn:addaccum4}
\frac{4(N_d+1)}{2^{\eta^{k_{baf}}}}
\end{equation}
where the multiplicative factor of 2 relative to Equation \eqref{eqn:addaccum3} comes from the denominator $\mu!$.

\subsubsection{Errors in per-cycle computation: Line 23 addition error}
\label{subsubsec:line23adderr}
The additions in Line 23 of Algorithm \ref{alg:divconq} initially contributes to the following maximum magnitude per each $g_{\theta_c}(t)$:
\begin{equation}
\label{eqn:addaccum5}
\frac{2N_{\mu}}{2^{\eta^{k_{baf}}}}
\end{equation}

\subsubsection{Errors in per-cycle computation: addition error and extrapolation error}
\label{subsubsec:divconqaddextcomberrr}
We now utilize the results in subsection \ref{subsec:err}. Equation \eqref{eqn:wsumintpnp2} just needs to be modified and adapted for analysis of extrapolation errors caused by addition error contributions. Since addition error analysis centers around contributions to $g_{\theta_c}(t) = \sum_{\omega} A_{\omega}e^{i\omega t}$, Equation \eqref{eqn:wsumintpnp2} needs to change to:
\begin{equation}
\label{eqn:errnoka}
\frac{1}{z_2}\sum_{m=z}^{z_2}\sum_{t=1}^{z_2}|w_{\theta_c,m}(t)|^2 < 5^2\sum_{\omega}|A_{\omega}|^2
\end{equation}

Let $e_{\ell,c,j}(t)$ refers to $w_{\theta_c,j}(t)$ when $g_{\theta_c}(t)$ is the Algorithm \ref{alg:divconq} Line $\ell$ addition error induced at time $t$, with $\ell$ referring to the line number. Let $e_{\ell,c,z-1}(t)$ refer to $g_{\theta_c}(t)$ itself.

Let us now discuss the Algorithm \ref{alg:divconq} Line 11 addition errors and their induced extrapolation errors. Given Equation \eqref{eqn:addaccum1}, $\sum_{\omega}|A_{\omega}|^2$ in Equation \eqref{eqn:errnoka} is bounded by:
\begin{equation}
\label{eqn:line11aomegasum}
\sum_{\omega}|A_{\omega}|^2 \leq \left(\frac{2N_f}{2^{\eta^{k_{baf}}}}\right)^2
\end{equation}
Therefore:
\begin{equation}
\label{eqn:line11percyclerrtot}
\frac{1}{z_2}\sum_{j=z-1}^{z_2} \sum_{t=1}^{z_2}|e_{11,c,j}(t)|^2 \leq 26\left(\frac{2N_f}{2^{\eta^{k_{baf}}}}\right)^2
\end{equation}

For the Algorithm \ref{alg:divconq} Line 18 addition error, $\sum_{\omega}|A_{\omega}|^2$ in Equation \eqref{eqn:errnoka} is bounded by, given Equation \eqref{eqn:addaccum4}:
\begin{equation}
\label{eqn:line18aomegasum}
\sum_{\omega}|A_{\omega}|^2 \leq \left(\frac{4(N_d+1)}{2^{\eta^{k_{baf}}}}\right)^2
\end{equation}
Therefore:
\begin{equation}
\label{eqn:line18percyclerrtot}
\frac{1}{z_2}\sum_{j=z-1}^{z_2} \sum_{t=1}^{z_2}|e_{18,c,j}(t)|^2 \leq 26\left(\frac{4(N_d+1)}{2^{\eta^{k_{baf}}}}\right)^2
\end{equation}

For the Algorithm \ref{alg:divconq} Line 23 addition error, $\sum_{\omega}|A_{\omega}|^2$ in Equation \eqref{eqn:errnoka} is bounded by, given Equation \eqref{eqn:addaccum5}:
\begin{equation}
\label{eqn:line23aomegasum}
\sum_{\omega}|A_{\omega}|^2 \leq \left(\frac{2N_{\mu}}{2^{\eta^{k_{baf}}}}\right)^2
\end{equation}
Therefore:
\begin{equation}
\label{eqn:line23percyclerrtot}
\frac{1}{z_2}\sum_{j=z-1}^{z_2} \sum_{t=1}^{z_2}|e_{23,c,j}(t)|^2 \leq 26\left(\frac{2N_{\mu}}{2^{\eta^{k_{baf}}}}\right)^2
\end{equation}

\subsection{Final error analysis}
\label{subsec:finalerr}
It is important to note that addition errors do not interact with each other in sense that each addition error analysis can be done separately in terms of an upper bound. Furthermore, pure extrapolation errors do not induce addition errors in sense, again, that addition error analysis can be done independently of pure extrapolation errors in terms of an upper bound - though addition errors do induce extrapolation errors. Therefore, per each cycle, there are pure extrapolation errors, addition errors and extrapolation errors induced by addition errors. This simplifies calculations.

Let us define $e_{\beta,\ell,a_1,a_2,..,a_{\eta^2}}(t)$ at cycle $\eta^2$. $\beta$ here refers to the error source algorithm. $\ell$ refers to the initial addition error source line number. (In case $\ell=0$, errors are pure extrapolation errors.) $a_j$ refers to `at which time $t'$ in cycle $c$' extrapolation error was induced, with $j$ referring to $j$th cycle, with the input source that induced the error having come from extrapolation error at time $a_k$ in cycle $k<j$. To summarize, extrapolation error function $e_{\beta,\ell,a_1,a_2,..,a_{\eta^2}}(t)$ at cycle $\eta^2$ was generated at time $t=a_{\eta^2}$ (therefore, this function equals zero at integer times $t<a_{\eta^2}$), with input coming from the extrapolation error function generated at time $a_{\eta^2-1}$ in cycle $\eta^2-1$, assuming that the function comes from the input of the extrapolation error function generated at time $a_{\eta^2-2}$ in cycle $\eta^2-2$ (note that time $a_{a_{\eta^2-2}}$ does not refer to time ordering in cycle $\eta^2-1$ or $\eta^2$ or $\eta^2-3$), assuming that the function comes from the input of the extrapolation error function generated at time $a_{\eta^2-3}$ in cycle $\eta^2-3$ and so forth. When $a_j=z-1$, the error function generated at time $a_{j+1}$ in cycle $j+1$ comes from the input of the extrapolation error function generated at time $a_{j-1}$ in cycle $j-1$. In case $a_j = z-1$ for all integer $1 \leq j \leq \eta^2$, then no extrapolation error was induced.

There are $\eta^2$ cycles - therefore, the sum of absolute value squared of extrapolation error functions caused by ideal $f_c(t)$ at the end of cycle $\eta^2$ (the final cycle) where $g_{\eta^2}(t)$ is calculated is bounded above by $q_{f_c}$ as, derived from Equation \eqref{eqn:wsumintpnp2}:

\begin{multline}
\label{eqn:pureext}
\frac{1}{z_2}\biggl[\sum_{a_{\eta^2}=z-1}^{z_2}..\sum_{a_1=z-1}^{z_2}\sum_{t=1}^{z_2}|e_{1,0,a_1,a_2,..,a_{\eta^2}}(t)|^2 - \\ \sum_{t=1}^{z_2}|e_{1,0,z-1,z-1,..,z-1}(t)|^2\biggr] \leq 2\eta^2 \left(\frac{5}{\eta^{\eta^{k_a}}}\right)^2\sum_{\omega}|A_{\omega}|^2
\end{multline}
\begin{equation}
\label{eqn:pureextextaa}
2\eta^2 \left(\frac{5}{\eta^{\eta^{k_a}}}\right)^2\sum_{\omega}|A_{\omega}|^2 \leq q_{f_c}
\end{equation}
with
\begin{equation}
\label{eqn:qfc}
q_{f_c} = 50\eta^2 (\eta^{\eta})^2 \left(\frac{1}{\eta^{\eta^{k_a}}}\right)^2
\end{equation}
with the assumption of $\sum_{\omega}|A_{\omega}| \leq \eta^{\eta}$ as justified from the definition of $f_c(t)$.

Contributions of Algorithm \ref{alg:algxt} addition errors - sum of absolute value squared of addition error functions plus absolute value squared of extrapolation error functions - at the end of cycle $\eta^2$, given Equation \eqref{eqn:adderrorxt} and \eqref{eqn:wsumintpnp2}, are bounded above by $q_{1,11}$ as:
\begin{equation}
\label{eqn:algxtaapre}
\frac{1}{z_2}\sum_{a_{\eta^2}=z-1}^{z_2}..\sum_{a_1=z-1}^{z_2}\sum_{t=1}^{z_2}|e_{1,11,a_1,a_2,..,a_{\eta^2}}(t)|^2 \leq q_{1,11}
\end{equation}

\begin{equation}
\label{eqn:algxterr}
q_{1,11} = \left[\frac{2\eta^{2\eta+1}}{2^{\eta^{k_{baf}}}}\right]^2 + 25\left[\frac{2\eta^{2\eta+1}}{2^{\eta^{k_{baf}}}}\right]^2 \left(\frac{1}{\eta^{\eta^{k_a}}}\right)^2
\end{equation}
Note the factor $25$. Algorithm \ref{alg:algxt} errors are produced only for cycle 1, since samples of $f(t)$ are only used for cycle 1 in Algorithm \ref{alg:divconq}.

Contributions of Algorithm \ref{alg:algxt} multiplication errors - sum of absolute value squared of addition error functions plus absolute value squared of extrapolation error functions - at the end of cycle $\eta^2$, given Equation \eqref{eqn:multerrorxt} and \eqref{eqn:wsumintpnp2}, are bounded above by $q_{1,14}$ as:
\begin{equation}
\label{eqn:algxtmulaapre}
\frac{1}{z_2}\sum_{a_{\eta^2}=z-1}^{z_2}..\sum_{a_1=z-1}^{z_2}\sum_{t=1}^{z_2}|e_{1,14,a_1,a_2,..,a_{\eta^2}}(t)|^2 \leq q_{1,14}
\end{equation}

\begin{equation}
\label{eqn:algxtmulerr}
q_{1,14} = \left[\frac{2\eta^{\eta+1}}{2^{\eta^{k_{bmf}}}}\right]^2 + 25\left[\frac{2\eta^{\eta+1}}{2^{\eta^{k_{bmf}}}}\right]^2 \left(\frac{1}{\eta^{\eta^{k_a}}}\right)^2
\end{equation}

When one translates $x(t)$ into $f(t)$ via $y(t)$, a sinusoid computation error is incurred. This contributes - sum of absolute value squared of addition error functions plus absolute value squared of extrapolation error functions - at the end of cycle $\eta^2$, given Equation \eqref{eqn:wsumintpnp2}:
\begin{equation}
\label{eqn:algxtfterpre}
\frac{1}{z_2}\sum_{a_{\eta^2}=z-1}^{z_2}..\sum_{a_1=z-1}^{z_2}\sum_{t=1}^{z_2}|e_{1,22,a_1,a_2,..,a_{\eta^2}}(t)|^2 \leq q_{1,22}
\end{equation}

\begin{equation}
\label{eqn:algxtfterr}
q_{1,22} = \left[\frac{2}{2^{\eta^{k_{bmf}}}}\right]^2 + 25\left[\frac{2}{2^{\eta^{k_{bmf}}}}\right]^2 \left(\frac{1}{\eta^{\eta^{k_a}}}\right)^2
\end{equation}
assuming $\eta^{\eta+1}/2^{\eta^{k_s}} \ll 2/2^{\eta^{k_{bmf}}}$.

From Equation \eqref{eqn:line11aomegasum} and \eqref{eqn:errnoka}, the Algorithm \ref{alg:divconq} Line 11 addition error contributions at the end of cycle $\eta^2$ - sum of absolute value squared of addition error functions plus absolute value squared of extrapolation error functions - can be summed up with upper bound $q_{2,11}$: 
\begin{equation}
\label{eqn:line11finalpurepre}
\frac{1}{z_2}\sum_{a_{\eta^2}=z-1}^{z_2}..\sum_{a_1=z-1}^{z_2}\sum_{t=1}^{z_2}|e_{2,11,a_1,a_2,..,a_{\eta^2}}(t)|^2 \leq q_{2,11}
\end{equation}

\begin{equation}
\label{eqn:line11finalpure}
q_{2,11} = 52\eta^2\left(\frac{2N_f}{2^{\eta^{k_{baf}}}}\right)^2
\end{equation}
where $\eta^2$ comes from the number of cycles, combined with Equation \eqref{eqn:wsumintpnp2} for the multiplicative factor of 2.

From Equation \eqref{eqn:line18aomegasum} and \eqref{eqn:errnoka}, the Algorithm \ref{alg:divconq} Line 18 addition error contributions at the end of cycle $\eta^2$ - sum of absolute value squared of addition error functions plus absolute value squared of extrapolation error functions - can be summed up with magnitude upper bound $q_{2,18}$:
\begin{equation}
\label{eqn:line18finalpurepre}
\frac{1}{z_2}\sum_{a_{\eta^2}=z-1}^{z_2}..\sum_{a_1=z-1}^{z_2}\sum_{t=1}^{z_2}|e_{2,18,a_1,a_2,..,a_{\eta^2}}(t)|^2 \leq q_{2,18}
\end{equation}

\begin{equation}
\label{eqn:line18finalpure}
q_{2,18} = 52\eta^2\left(\frac{4(N_d+1)}{2^{\eta^{k_{baf}}}}\right)^2
\end{equation}

From Equation \eqref{eqn:line23aomegasum} and \eqref{eqn:errnoka}, the Algorithm \ref{alg:divconq} Line 23 addition error contributions at the end of cycle $\eta^2$ - sum of absolute value squared of addition error functions plus absolute value squared of extrapolation error functions - can be summed up with magnitude upper bound $q_{2,23}$:
\begin{equation}
\label{eqn:line23finalpurepre}
\frac{1}{z_2}\sum_{a_{\eta^2}=z-1}^{z_2}..\sum_{a_1=z-1}^{z_2}\sum_{t=1}^{z_2}|e_{2,23,a_1,a_2,..,a_{\eta^2}}(t)|^2 \leq q_{2,23}
\end{equation}

\begin{equation}
\label{eqn:line23finalpure}
q_{2,23} = 52\eta^2\left(\frac{2N_{\mu}}{2^{\eta^{k_{baf}}}}\right)^2
\end{equation}

Therefore, the upper bound for sum of absolute value squared of error functions is:

\begin{multline}
\label{eqn:nondecayz}
\frac{1}{z_2}\sum_{\beta}\sum_{\ell}\sum_{a_{\eta^2}=z-1}^{z_2}..\sum_{a_1=z-1}^{z_2}\sum_{t=1}^{z_2}|e_{\beta,\ell,a_1,a_2,..,a_{\eta^2}}(t)|^2 \leq \\ q_{f_c} + q_{1,11} + q_{1,14} + q_{1,22} + q_{2,11} + q_{2,18} + q_{2,23}
\end{multline}

This demonstrates that error contributions can be made negligible with right selections for the constants notated with $k$.

\subsection{Constants k}
\label{subsec:constantsk}
All constants with $k$ label are positive integers.
\begin{itemize}
    \item $k_a = k_d$: used in post-cutoff minimum attenuation $a_m$ ($k_a$) for filter $H_f$ and in maximum error magnitude for angular frequency $|\omega|\leq \omega_c$ for derivative filter $H_{d,\mu}$. Set as: $k_a=k_d=2$. This is about the design choice to attenuate a high frequency signal with amplitude magnitude of $\eta^{\eta}$ sufficiently as to not hamper with the computation of $n_h$, the number of undirected Hamiltonian paths in a graph. 
    \item $k_c$: used in cut-off frequency $\omega_c$ of both filter $H_f$ and derivative filters $H_{d,\mu}$. $k_c = 1$.
    \item $\eta^{k_{bai}}=\eta^{k_{bmi}}$: number of integer bits kept in product computation ($\eta^{k_{bmi}}$) or in sum computation ($\eta^{k_{bai}}$). Via Equation \eqref{eqn:kbaikbmi}, related to $k_a$ and $k_c$.
    \item $k_{ffc}$: used in time complexity of finding filter coefficients. Related to $k_a$, $k_c$ and $k_p$ via Equation \eqref{eqn:kffc}.
    \item $k_p$. $1/\eta^{\eta^{k_p}}$ refers to maximum error magnitude when computing each filter coefficient. $k_p =7$, given consideration in Equation \eqref{eqn:fcerrmulerr}, assuming $k_{bmf}=5$.
    \item $\eta^{k_{baf}}$, $\eta^{k_{bmf}}$: the number of fractional digits kept in sum computation ($\eta^{k_{baf}}$), in product computation ($\eta^{k_{bmf}}$). Set as: $k_{baf} = 3$, $k_{bmf}=5$, with the reason for this setup given by Equation \eqref{eqn:mulerradderr1}.
    \item $k_s$: $1/\eta^{\eta^{k_s}}$ represents maximum magnitude of error in computing $e^{i\omega t}$. Set as: $k_s=7$, coming from considering Equation \eqref{eqn:sinvsadd} and \eqref{eqn:sinvsmult}.
    \item $k_{aa}$, $k_{ca}$, $k_{ab}$, $k_{cb}$, $k_{ac}$, $k_{cc}$, $k_{pc}$: undetermined constant coefficients used in Equation \eqref{eqn:filterordernfnd} in determining filter order ($k_{aa},k_{ca}$), in Equation \eqref{eqn:kbaikbmi} in determining $k_{bai}$ ($k_{ab},k_{cb}$), in Equation \eqref{eqn:ffctimecomplexity} in reflecting filter coefficient precision, $k_a$ and $k_c$ dependence on filter coefficient time complexity ($k_{ac}$, $k_{cc}$, $k_{pc}$).
    \item $k_{\mu}$: used in $N_{\mu} = \eta^{k_{\mu}}$, which is the final order of numerical derivative computed. $N_f,N_d \gg N_{\mu}$ is required. Thus, because of Equation \eqref{eqn:filterordernfnd}, $k_a k_{aa} + k_c k_{ca} \gg k_{\mu}$. $k_{\mu}=2$.
\end{itemize}

Let us think of relationships between different $N$s and constant $k$s. As aforementioned, $N_f,N_d \gg N_{\mu}$. $N_d$ is determined by $N_{\mu}$ (and thus $k_{\mu}$), $k_a$ and $k_c$. $N_f$ is determined by $k_a$ and $k_c$ (along with $k_{aa}$ and $k_{ca}$ for both $N_f$ and $N_d$, though they are not parameters we can choose - and $\eta^{k_c} = \omega_c$). This suggests that the real choice we have is about $k_a$ and $k_c$. 

\subsection{Time complexity analysis}
\label{subsec:timecomplexity}
In Algorithm \ref{alg:algxt}, the line that dominates for time complexity is Line 14. Time complexity of Algorithm \ref{alg:algxt} is thus given by $O(\eta^2 c_{sin})$, where $c_{sin}$ refers to time complexity of calculating each $e^{i\omega t}$ to precision set by $k_s$. In Algorithm \ref{alg:divconq}, the dominant line in terms of time complexity is Line 18 with $O(\eta^2(N_f+N_d)N_{\mu}N_d c_{mult})$, with $C_{mult}$ refering to time complexity of multiplication of two numbers, with each number constrained to have at most $(\eta^{k_{bmi}}+\eta^{k_{bmf}})/2$ binary digits. Rest of time complexity involves finding filter coefficients, with assumed time complexity $O(\eta^{k_{ffc}})$. Out of these three factors, $O(\eta^{k_{ffc}})$ would dominate - therefore, whole computation has time complexity of $O(\eta^{k_{ffc}})$.

\section{Conclusion}
\label{sec:conclusion}
The undirected Hamiltonian path problem was reduced to a signal filtering problem based on a special graph encoding into $f(t)$. Construction of $f(t)$ assigns distinct and unique angular frequency for Hamiltonian paths - zero frequency - with each possible path contributing amplitude of $1$. Non-zero frequencies with potential non-zero amplitude are in the domain $1/|V|^{|V|+1} \leq |\omega| \leq 1$ for $f(t)$. This makes the lowpass filtering problem of extracting the zero-frequency amplitude - the number of Hamiltonian paths - very difficult if one sticks with the conventional filtering strategies, given wide bandwidth to be filtered out.

Assuming validity of the filter order bound and the filter time complexity assumption corresponding to required filter design, one can find the way to filter out wide bandwidth and extract the zero-frequency signal efficiently, taking $O(|V|^{k_{ffc}})$, where $k_{ffc}$ is constant but is left undetermined, as this depends on exact time complexity of finding filter coefficients.

The lowpass wide bandwidth filtering strategy was developed based on a frequency `divide and conquer' idea. One must filter out angular frequency $1/2$ to $1$ and then $1/4$ to $1/2$ and so on until one reaches $1/{\eta}^{\eta+1}$, not requiring change in cutoff frequency, accounting for time re-scaling. Each frequency range filtering is then considered to be a filtering cycle, or a filtering - extrapolation cycle. In order to provide sufficient number of filter input samples at the beginning of a subsequent cycle, filter output extrapolation is necessary at each cycle. This is done through careful construction of a local polynomial based on numerical differentiation.


\end{document}